# Scenario of Use Scheme: Threat Model Specification for Speaker Privacy Protection in the Medical Domain

*Mehtab Ur Rahman, Martha Larson, Louis ten Bosch, Cristian Tejedor-García*

Center for Language Studies
Radboud University Nijmegen, the Netherlands

`(mehtab.rahman, martha.larson, louis.tenbosch, cristian.tejedorgarcia)@ru.nl`

## Abstract

Speech recordings are being more frequently used to detect and monitor disease, leading to privacy concerns. Beyond cryptography, protection of speech can be addressed by approaches, such as perturbation, disentanglement, and re-synthesis, that eliminate sensitive information of the speaker, leaving the information necessary for medical analysis purposes. In order for such privacy protective approaches to be developed, clear and systematic specifications of assumptions concerning medical settings and the needs of medical professionals are necessary. In this paper, we propose a *Scenario of Use Scheme* that incorporates an Attacker Model, which characterizes the adversary against whom the speaker's privacy must be defended, and a Protector Model, which specifies the defense. We discuss the connection of the scheme with previous work on speech privacy. Finally, we present a concrete example of a specified Scenario of Use and a set of experiments about protecting speaker data against gender inference attacks while maintaining utility for Parkinson's detection.

## 1. Introduction

Advancements in artificial intelligence have enabled the use of speech processing in medical applications, like the early detection and classification of neurodegenerative diseases such as Parkinson's disease [1, 2] and Alzheimer's disease [3, 4]. The growing use of speech data in the medical domain raises privacy concerns, since the human voice carries sensitive personal information. Besides health status, our voices convey information about gender, emotions, age, and identity, which can potentially allow unwanted inference of sensitive attributes or unwanted speaker identification [5, 6].

The widely-known *security triad* consists of Confidentiality, Integrity, and Accessibility. This triad also applies to privacy, i.e., protecting people's sensitive information. In the medical domain, protecting Integrity can mean protecting the medically relevant information in the data and protecting an intact version of the original recording is not necessary. For this reason, recent work on perturbation, disentanglement, and re-synthesis, which modifies either the original audio or stored representations to remove non-essential information, is promising for use in medical settings [7, 8, 9].

However, these techniques do not provide a one-size-fits all solution for speech privacy in the medical domain. Instead, the privacy techniques that are needed depend on the particular setting or scenario in which voice data is used. Correctly characterizing application scenarios requires close communication between computer scientists, data managers, and medical professionals. In this paper, we introduce a *Scenario of Use Scheme* whose aim is to make the descriptions of speech privacy protection scenarios explicit and systematic across research studies. The scheme is a format in which an extended threat model can be specified, including specifications for both an Attacker Model and a Protector Model. The scheme also reduces potential confusion about which research results can be directly compared or the situations in which results are applicable.

The remainder of the paper is structured as follows. In Section 2, we introduce our Scenario of Use Scheme. Then, in Section 3, we discuss how this scheme emerges from existing work and how it can help to improve the current practice of threat modelling. Finally, in Section 4, we provide an example of a specification of a Scenario of Use and a simple study on medical speech data that uses this specification. We hope our paper encourages discussion of standardization of threat modelling and uptake of systematic specifications in research papers.

## 2. Scenario of Use Scheme

The development of security and privacy defenses starts with the specification of a threat model, which describes the attacks that are expected. A threat model contains an characterization of the attacker, including the goal that the attacker is trying to achieve, the vulnerability that provides the attacker with the opportunity to achieve that goal, and the additional resources at the attackers disposal [10].

Conventionally, the protector is assumed to be the mirror image of the attacker, i.e., the protector's objective is to block the attacker's objective. However, as we move to protecting speaker's privacy in the medical domain, the description of the protector must be more sophisticated. The goal of the protector is to block the attack, but to do so in a manner that does not at the same time destroy the usefulness of the data for medical analysis. Also, the protector might have limited knowledge of the attacker and/or possible countermeasures might be limited. For these reasons, we propose extending the traditional specification of a threat model, which only describes the attacker, to a Scenario of Use format, which describes both attacker and protector. Note that the Scenario of Use Scheme may ultimately be helpful in supporting systematic communication with stakeholders in other application domains beyond the medical domain, however, in this work our motivation derives from the need to define the privacy problem in the medical domain.

The Scenario of Use Scheme is presented in Table 1. This scheme sets out the dimensions that must be specified in order to start developing privacy protection within a medical domain. Before research starts, we recommend that privacy experts, data managers, and medical experts come together to specify the individual dimensions of the scheme for the threats to be studied.

Table 1: *Our proposed Scenario of Use Scheme for systematic description of speech privacy applications*

| | **Attacker Model (Attack on Privacy)** | | **Protector Model (Protection of Privacy)** |
|---|---|---|---|
| Objective | The information about the target speakers the attacker seeks to obtain. | Objective | **Defense objective**: The information in the spoken audio that must be protected.<br><br>**Utility objective**: The information in the spoken audio that must be retained. |
| Opportunity | The access of the attacker to the target, including availability of target speakers' audio. Also, access to (and any knowledge about) the protection. | Opportunity | The possibilities available for protection (i.e., countermeasures). Also, access to (and any knowledge about) the attack. |
| Additional Resources | The knowledge, data, and compute available to the attacker to carry out the attack (beyond the access to the target specified under Opportunity). | Additional Resources | The knowledge, data, and compute available to the protector to defend against attack (beyond the access to the attack specified under Opportunity). |

Here, we mention how the Scenario of Use Scheme supports discussion of underlying assumptions with stakeholders and ensures a solid, transparent foundation for research papers. In medical research, the dimension *Objective* might differ among use cases. For instance, in a specific use case, speakers may remain identifiable if their audio is associated with their medical record. However, it is not necessary to retain other information, such as emotional state. In this case, the *Objective* would specify the attributes that the attacker seeks to infer and the protector seeks to protect. The dimension *Opportunity* explicitly describes what the attacker knows about the protection and what the protector knows about the attack, disambiguating potential confusing terms like black box, gray box, and white box. The dimension *Additional Resources* specifies knowledge, data, and compute that can be leveraged beyond the specific setting of the attack. In the next section, we discuss how differences in the existing work can be systematically captured using this Scenario of Use Scheme. If a specific system is being studied, Opportunity would express system-engineering considerations. However, here we focus on research where data protection is studied independently of a specific system.

## 3. Relation of Scheme to Existing Work

The dimensions of our Scenario of Use Scheme in Table 1 are inspired by the seminal work on threat modelling, e.g., [10], as well as existing literature on voice privacy. Most dimensions can be identified as already in use in the literature. In this section, we discuss the connection of the Scenario of Use Scheme to previous work by examining three examples (cf. Table 2). The examples illustrate the ability of our Scenario of Use Scheme to make the settings of the dimensions explicit and express them in a standard way across papers. Our aim is that, in papers using the scheme, foundational assumptions will be readily identifiable, straightforward to compare with other papers, and easier to discuss (e.g., with medical professionals).

We start with the VoicePrivacy Challenge [11, 14], because it the leading example of a benchmark, whose purpose is to drive forward the state of the art by supporting fair comparison of privacy solutions for voice data. The 2024 VoicePrivacy overview paper [11] expresses the dimensions of the Scenario of Use Scheme in the sections that provide the task definition, the attack model, the training resources and the privacy and utility metrics. Because the VoicePrivacy Challenge is so well specified, expressing it as a Scenario of Use Specification following the dimensions of our Scenario of Use Scheme is essentially a translation (see left column of Table 2).

Two other papers that develop voice privacy techniques and discuss threat models are [12] and [13]. These papers both discuss threat modelling and specifications for protection, but less systematically. We have located the relevant information in the papers re-expressed as a Scenario of Use Specification (middle column of Table 2 is [12] and right column is [13]).

Considering the three Scenario of Use Specifications in Table 2 side by side demonstrates how standardization of the threat modelling makes it easier to identify differences between different papers. For instance, while the VoicePriacy Challenge 2024 [11] focuses on the case that the attacker has audio of the target speaker, in [12] the assumption is that the attacker has features that have been extracted from the audio of the target speaker. Also, the VoicePrivacy Challenge 2024 [11] assumes that the attacker has access to the voice anonymization system that is used for protection, in [13] the attacker attacks protected utterances without knowledge of the nature of the protection. In the medical domain, it is crucial to identify and discuss differences with domain experts. For instance, some applications need to protect speech before it is analyzed or in real time. These distinctions must be clear so that they can be discussed during the research and development process.

The Scenario of Use Scheme could also help authors to carefully distinguish between the modelling phase and the implementation phase that designs the experimental setup and carries out the experiments. The modelling phase, as with traditional threat modelling, is conceptual and the experiment phase is practical, and may only approximate the conceptual problem. For instance, in [12] the distinction is blurred when the paper refers to "training the threat model".

## 4. Example of the Scheme in Action

We study the case of protecting speaker data from a gender inference attack. We assume that the protector has knowledge of the attacker and can use the attack classifier to create perturbations to protect the data using the well-known Projected Gradient Descent (PGD) approach [15]. The protector's Defense Objective is to protect the gender attribute of the speech data, while the Utility Objective is to maintain the classification accuracy of the diagnosis classifier. We specify the Scenario of Use explicitly in Table 3. Our proposal in this paper is that all speech privacy research should include such a specification table. Although this study is simple and straightforward, it is interesting because it illustrates the Scenario of Use Scheme in action and is also the first time, to our knowledge, that PGD has been tested to protect Parkinson data.

### 4.1. Experimental Setup

#### 4.1.1. Dataset

In this study, we used a subset of the openly-available NeuroVoz corpus [16, 17], which contains audio recordings of 108 partic-ipants involved in various speech tasks. The recordings were collected by Universidad Politécnica de Madrid and the Departments of Otorhinolaryngology and Neurology of HGUGM (Spain). All of the recordings in this corpus were sampled at a frequency of 44.1 kHz. We selected a subset of 1351 recordings for our experiment. A total of 1141 recordings were used for training, consisting of 530 HC and 611 PD, which included five vowels and a BARBAS sentence. This training set comprised 656 male and 485 female speakers. The test set consisted of 210 recordings, with 108 HC and 102 PD, which included the ACAMPADA and ABLANDADA sentences. This test set comprised 119 male and 91 female speakers. Table 4 overviews the dataset, breaking down PD, HC, and gender.

Table 2: *Scenario of Use Specifications in existing work. Most papers include statements of the assumptions made about the attacker and the protector. However, this information is not provided in a standardized format, it is often implicit, and somethings it is incomplete. To create this table, we took information from three papers and re-expressed it in terms of our Scenario of Use. Where the information is not clear from the paper, we provide the best interpretation that we could.*

| | **VoicePrivacy 2024 Challenge** [11] | **Variational Information Feature Extraction** [12] | **Gender-ambiguous voices** [13] |
|---|---|---|---|
| **Attacker Model (Attack on Privacy)** | | | |
| Objective | Attacker attempts to identify the speaker. | Attacker attempts to identify or verify the speaker. | Attacker attempts to acquire the gender and identity of the speaker. |
| Opportunity | The attacker has access to anonymized target audio, to enrollment audio for each possible speaker, and to the voice anonymization that was used by the protector. | The attacker has access to anonymized target features, to enrollment audio for each possible speaker, and to the same feature extractor as the protector. | Attacker has access to protected target audio. |
| Additional Resources | Attacker can train a verification system. | Attacker can train a speaker identification and verification system. | Attacker has a gender classifier and a speaker verification system. |
| **Protector Model (Protection of Privacy)** | | | |
| Objective | Defense objective: Defend the speaker's identity. Utility objectives: Maintain linguistic content and emotional states. | Defense objective: Defend the speaker's identity. Utility objectives: Sound classification performance. | Defense objective: Defend the speaker's gender and identity. Utility objectives: Speech recognition. |
| Opportunity | Protector uses a voice anonymization system. | Protector can modify the feature extractor (variational feature extraction). | Protector has a GenGAN speech transformation/synthesis network. |
| Additional Resources | Baseline anonymization systems and additional training data. | — | — |

Table 3: *Scenario of Use Specification for our example case of gender privacy protection.*

| | **Attacker Model (Attack on Privacy)** | | **Protector Model (Protection of Privacy)** |
|---|---|---|---|
| Objective | The attacker aims to acquire the gender attribute of the target speakers. | Objective | **Defense objective:** The protector aims to keep the gender attribute of the speech data concealed. **Utility objectives:** The protector aims to maintain the classification accuracy of the diagnosis classifier on the speech data. |
| Opportunity | Attacker has speech data from target speakers, which is not labeled with gender. | Opportunity | Protector modifies the data of the target speakers. Protector has a copy of the trained gender classifier used for the attack (used for perturbation). |
| Additional Resources | The attacker has access to labeled speaker data to train a gender classifier. The training data is drawn from the same distribution as the target user data. Also, the training data is the same data that was used to train the diagnosis classifier. | Additional Resources | The protector does not use additional resources. |

### 4.1.2. Data Preparation

In our study, each audio signal is normalized by dividing each sample by the maximum absolute amplitude value within the signal. As the length of audio signals varies, we segmented them into fixed-length chunks of 0.4 second long with 50% overlap between consecutive segments. A Mel-scale spectrogram transformation was applied to each audio recording, converting the signal into the time-frequency domain. Mel spectrograms were computed by applying the Short-Time Fourier Transform (STFT) with a 15-millisecond Hanning window. The hop length set to the half the window length resulted in 50% overlap between consecutive window segments. The Mel spectrograms are converted to decibel scale, followed by standard normalization to ensure all features have a zero mean and unit standard deviation. These preprocessing steps were the same for both the diagnosis and gender classifiers.

Table 4: *Dataset overview*

| | | HC | PD | Total |
|---|---|---|---|---|
| Diagnosis | Train | 530 | 611 | 1141 |
| Classifier | Test | 108 | 102 | 210 |
| | | **M** | **F** | |
| Gender | Train | 656 | 485 | 1141 |
| Classifier | Test | 119 | 91 | 210 |

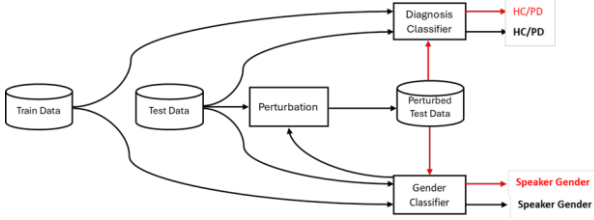

Figure 1: *Example Gender Protection Evaluation Pipeline*

#### 4.1.3. Gender Classifier and Protective Perturbations

Figure 1 illustrates our example of the gender protection pipeline. We used a Convolutional Neural Network (CNN) for gender classification which employs two convolutional blocks with $3 \times 3$ kernels, stride $1$ and padding $1$. Each convolutional block is followed by batch normalization, ReLU activation, $2 \times 2$ max pooling and $0.2$ dropout. The feature maps are then passed through global average pooling and flattened. Finally, a Multilayer Perceptron (MLP) with ReLU activation is used, followed by a softmax layer for classification. The model was trained using a batch size of $32$ and $30$ epochs. For the optimization process, we employed the cross-entropy loss function and the Adam optimizer with a learning rate of $0.001$.

For gender protection, adversarial speech samples from the test set were generated using PGD [15] applied to our trained gender classifier. Goodfellow et al. [18] introduced the Fast Gradient Sign Method (FGSM), a single-step technique that employs the input's gradient with respect to the loss function to craft adversarial examples with a high probability of fooling the model. PGD is an iterative extension of FGSM, where the perturbation values are small and constrained within a specific range. The update rule for PGD is defined as:

$$x_{t+1} = \Pi_{x+S} \left( x_t + \alpha \cdot \text{sgn} \left( \nabla_x L(\theta, x_t, y) \right) \right) \quad (1)$$

where $x_t$ is the adversarial example at iteration $t$, $\Pi_{x+S}$ denotes the projection that ensures the perturbed samples remain within the valid range, $\alpha$ is the step size, $\nabla_x L(\theta, x, y)$ is the gradient of the loss function with respect to the input $x_t$, $\theta$ are the model parameters, and $y$ is the true label. We used perturbations of size $\epsilon = 0.1$, $\alpha = 0.0005$ and $20$ iterations.

#### 4.1.4. Diagnosis Classifier

For the diagnosis classifier, we utilized a CNN model consisting of three convolutional blocks, each comprising a $3 \times 3$ kernel size, stride of $1$ and padding of $1$. Batch normalization, ReLU activation, $2 \times 2$ max pooling and $0.2$ dropout were applied after each convolutional layer to enhance the network's performance and prevent overfitting. The fully connected layers consist of one hidden layer with $128$ neurons and a ReLU activation function, followed by an output layer with the number of classes as the output size. During training, we used a batch size of $64$ and trained the model for $50$ epochs. The optimization process employed the cross-entropy loss function and the Adam optimizer with a learning rate of $0.001$.

Table 5: *Gender classifier performance on original and perturbed speech data. ACC (Accuracy), AUC (Area under the ROC Curve), R (Recall), P (Precision), F (Female), M (Male)*

|  | ACC | AUC | F1 | R(M) | P(M) | F1 (M) | R(F) | P(F) | F1(F) |
|---|---|---|---|---|---|---|---|---|---|
| Original data | 87.66 | 0.88 | 0.88 | 0.93 | 0.81 | 0.87 | 0.83 | 0.95 | 0.88 |
| Perturbed data | 57.50 | 0.51 | 0.82 | 0.03 | 0.85 | 0.06 | 1.00 | 0.57 | 0.73 |

Table 6: *Diagnosis classifier performance on original and perturbed speech data. ACC (Accuracy), AUC (Area under the ROC Curve), R (Recall), P (Precision), HC (Healthy Control), PD (Parkinson's Disease)*

|  | ACC | AUC | F1 | R(HC) | P(HC) | F1 (HC) | R(PD) | P(PD) | F1(PD) |
|---|---|---|---|---|---|---|---|---|---|
| Original data | 87.66 | 0.88 | 0.86 | 0.90 | 0.88 | 0.89 | 0.85 | 0.88 | 0.86 |
| Perturbed data | 80.97 | 0.80 | 0.80 | 0.89 | 0.79 | 0.84 | 0.71 | 0.85 | 0.77 |

### 4.2. Results

Table 5 presents the performance metrics for the gender classifier on both the original and perturbed speech data. Protected data is considered to improve privacy if it significantly reduces the prediction performance of the gender classifier. Ideally, the Area Under the Curve (AUC) should reach 0.50, the level of random guessing. Here, we see that the the AUC drops to from 0.88 to 0.51, which indicates successful protection since the performance of the attack classifier is nearly random on the protected data.

Table 6 presents the performance metrics for the diagnosis classifier. After the speech data has been protected, the accuracy drops compared to the performance of the diagnosis classifier on the original data. However, this drop is relatively small: 6.69%. In future work, we will focus on developing techniques specifically designed to maintain the utility of the speech data for medical analysis.

## 5. Conclusion and Outlook

In this paper, we have proposed a Scenario of Use Scheme including Attacker Model and Protector Model specifications. The Scheme can be used to systematically describe application scenarios and support communication between developers and medical practitioners. Communication is essential to ensure that threat/protection models are as realistic as possible and to help prioritize which attributes and information need protection.

We advocate that papers and projects that propose or develop speech privacy techniques explicitly specify the Scenario of Use Scheme in order to promote clarity and comparability. Looking forward, we anticipate that the Scenario of Use Scheme will support the 'data bookkeeping' necessary to design valid speech privacy evaluation pipelines. These pipelines require mapping threat models onto data sets and data set splits, and this mapping must be done carefully to avoid unintended mixing of attacker and protector resources. Some threat models are harder to study because we lack the sufficient or sufficiently diverse data. We also anticipate that Scenario of Use Scheme will help to make clear both to speech privacy researchers and medical practitioners where additional data is needed.

## 6. Acknowledgements

This work was supported by the NWO research programme NGF AiNed Fellowship Grants under the project Responsible AI for Voice Diagnostics (RAIVD) - NGF.1607.22.013.